\documentclass{article} 
\usepackage{iclr2025_conference,times}

\usepackage{amsmath,amsfonts,bm}

\def\eqref#1{equation~\ref{#1}}

\def\1{\bm{1}}

\DeclareMathAlphabet{\mathsfit}{\encodingdefault}{\sfdefault}{m}{sl}
\SetMathAlphabet{\mathsfit}{bold}{\encodingdefault}{\sfdefault}{bx}{n}

\usepackage{hyperref}
\usepackage{url}
\usepackage{graphicx}
\usepackage{caption}
\usepackage{subcaption}
\usepackage{amsmath,amssymb}
\usepackage{listings}
\usepackage{xcolor}
\usepackage{placeins}

\definecolor{lightgray}{gray}{0.95}
\iclrfinalcopy

\title{EvoGit: Decentralized Code Evolution via Git-Based Multi-Agent Collaboration}

\author{Beichen Huang, Ran Cheng, Kay Chen Tan \\
Department of Data Science and Artificial Intelligence\\
The Hong Kong Polytechnic University\\
}

\begin{document}

\maketitle
{
\renewcommand\thefootnote{}
\footnotetext{
Contact Emails: \texttt{beichen.huang@connect.polyu.hk}, \texttt{ranchengcn@gmail.com},\\\texttt{kaychen.tan@polyu.edu.hk}
}
\addtocounter{footnote}{0}
}

\begin{abstract}
We introduce {EvoGit}, a decentralized multi-agent framework for collaborative software development driven by autonomous code evolution. EvoGit deploys a population of independent coding agents, each proposing edits to a shared codebase without centralized coordination, explicit message passing, or shared memory. 
Instead, all coordination emerges through a Git-based \emph{phylogenetic graph} that tracks the full version lineage and enables agents to asynchronously read from and write to the evolving code repository.
This graph-based structure supports fine-grained branching, implicit concurrency, and scalable agent interaction while preserving a consistent historical record. Human involvement is minimal but strategic: users define high-level goals, periodically review the graph, and provide lightweight feedback to promote promising directions or prune unproductive ones.
Experiments demonstrate EvoGit’s ability to autonomously produce functional and modular software artifacts across two real-world tasks: (1) building a web application from scratch using modern frameworks, and (2) constructing a meta-level system that evolves its own language-model-guided solver for the bin-packing optimization problem.
Our results underscore EvoGit’s potential to establish a new paradigm for decentralized, automated, and continual software development.
EvoGit is open-sourced at \href{https://github.com/BillHuang2001/evogit}{\texttt{github.com/BillHuang2001/evogit}}.
\end{abstract}

\section{Introduction}

Large language models (LLMs) have significantly advanced the automation of software development, excelling at tasks such as code generation, debugging, and documentation~\cite{zhao_survey_2023,minaee_large_2024,yang_qwen2_2024,gemini_team_gemini_2023,openai_gpt-4_2024}. Building upon this foundation, recent efforts have embedded LLMs into autonomous coding agents~\cite{chatdev_2023,yang_swe-agent_2024,hong2024metagpt,islam2024mapcoder}, enabling the execution of multi-step development workflows through tool usage, memory, and interaction policies~\cite{yao2023react,xi2023rise,wang_survey_2024}.

While promising, current frameworks face critical limitations. Most rely on scalar reward signals, ground-truth unit tests, or dense human feedback to supervise agent behavior. These assumptions are ill-suited for open-ended software engineering, where success criteria are emergent and multifaceted. Furthermore, agent collaboration is typically centralized and synchronous, requiring a global orchestrator and shared memory, which restricts scalability and robustness. Critically, the development process often lacks traceability: as code branches diverge and recombine, it becomes increasingly difficult to understand the provenance of design decisions or to reproduce successful outcomes.

To overcome these limitations, we present {EvoGit}—a decentralized, Git-native framework for collaborative software evolution. EvoGit models software development as an asynchronous, multi-agent search over a version space. Each autonomous agent independently proposes mutations or crossovers of existing code, and all code versions are stored as nodes in a directed acyclic graph (DAG) maintained through Git. Edges denote valid semantic transitions between versions, capturing evolutionary lineage without enforcing global synchronization or communication.

Rather than relying on reward functions or predefined tests, EvoGit organizes development around structural traceability and partial-order semantics. This version graph acts as both the memory and substrate of interaction: agents read from it, evolve code through local transformations, and write back new branches, enabling highly scalable parallel exploration. Human developers remain in the loop only at key moments, such as initialization, periodic inspection, and minimal high-level feedback. Key features of EvoGit include:
\begin{itemize}
    \item \textbf{Decentralized evolution:} Agents evolve code independently, coordinating solely through the shared version graph.
    \item \textbf{Reward-free development:} Progress is driven by structural transformations and lineage growth, not scalar optimization.
    \item \textbf{Lineage-aware crossover:} Merges between branches preserve semantic context and inheritance relationships.
    \item \textbf{End-to-end traceability:} Every edit is recorded as a commit, enabling inspection, reproducibility, and rollback.
\end{itemize}

To demonstrate EvoGit's effectiveness, we evaluate it on two real-world tasks. The first task involves evolving a fully interactive single-page web application from an empty scaffold, testing agents' ability to coordinate UI layout, modular structure, and logic without any visual feedback. The second task benchmarks EvoGit’s support for meta-level development by constructing an automated algorithm design pipeline, i.e., a system that evolves its own LLM-guided solver for the bin-packing problem. Together, these experiments highlight EvoGit's capacity for both user-facing application development and self-adaptive meta-coding.

\begin{figure}[htpb]
    \centering
    \includegraphics[width=\linewidth]{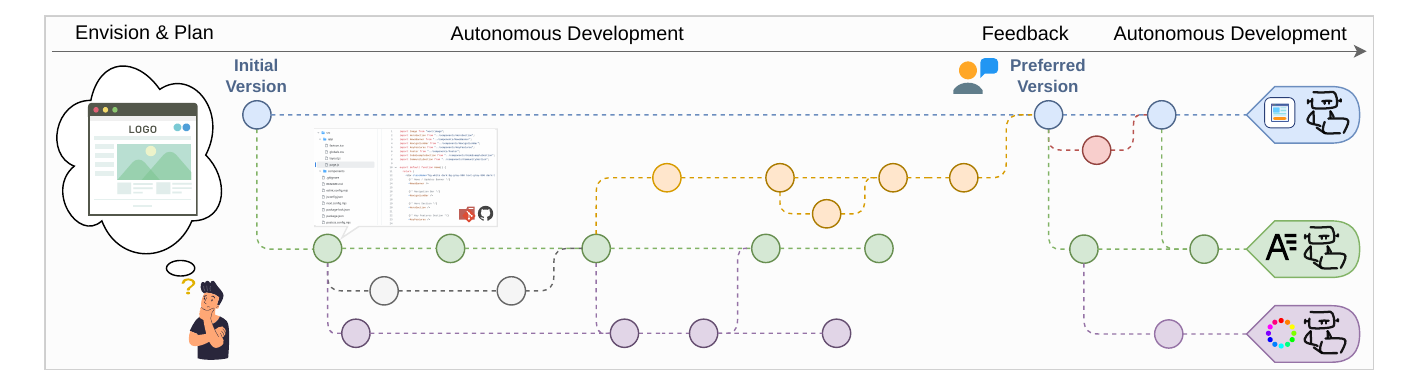}
\caption{
The {EvoGit} workflow. A human initializes the system by defining high-level goals and providing a seed codebase. Multiple autonomous coding agents independently explore, mutate, and recombine software artifacts on separate branches. All versions are recorded in a shared Git-based version graph, which serves as the coordination substrate. The human can periodically inspect the graph, select promising branches, and provide minimal feedback to guide further evolution.
}
    \label{fig:evogit-workflow}
\end{figure}

An overview of the EvoGit architecture is shown in Figure~\ref{fig:evogit-workflow}. By treating code as a population and development as an open-ended evolutionary process, EvoGit offers a scalable and inspectable paradigm for autonomous software creation.

\section{Methodology}

\begin{figure}[htpb]
    \centering
    \includegraphics[width=\linewidth]{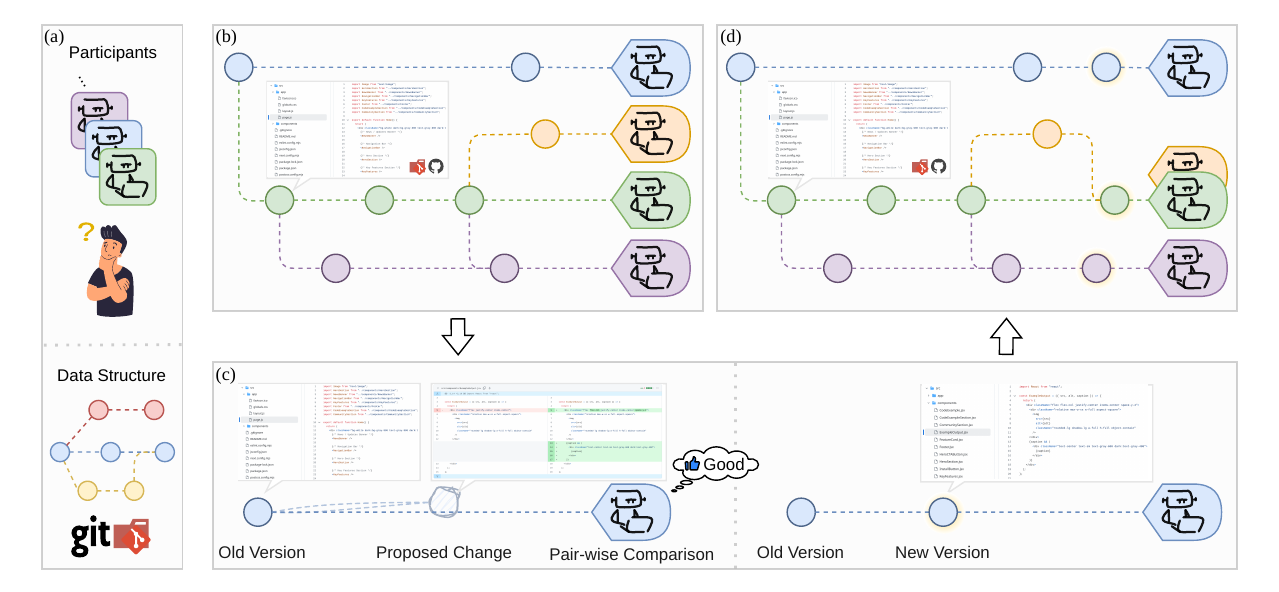}
    \caption{
    Overview of the EvoGit methodology. (a) EvoGit comprises two types of participants: a population of autonomous coding agents and a human product manager. The core of the system is a Git-backed phylogenetic graph that captures the complete history of code evolution. (b) The graph is composed of nodes representing code versions; agents operate on a dynamic subset of the current best versions. (c) Each agent proposes a minor code modification, which is locally evaluated using a pairwise comparison mechanism. If accepted, the new version is added as a node in the graph, and the agent continues from that point. (d) EvoGit supports two primary operations: mutation (one parent, one child) and crossover (two parents, one child), enabling both fine-grained updates and structural recombination across divergent branches.
    }
    \label{fig:method}
\end{figure}

\subsection{Core Components}

EvoGit conceptualizes software development as a multi-agent evolutionary process grounded in a shared and decentralized structure: \textbf{phylogenetic graph}, which records the full lineage of all code versions. The system is composed of four core components:

\begin{itemize}
    \item \textbf{Autonomous Agents:} Each agent operates independently and is powered by a large language model. Agents navigate the version space by proposing minor code edits (\emph{mutations}) or combining different versions via structured \emph{crossovers}, thereby contributing new nodes to the evolving graph.

    \item \textbf{Human Product Manager:} A human oversees the development process by defining the initial design intent and periodically interacting with the graph to select promising branches, prune ineffective ones, or provide sparse, high-level guidance.

    \item \textbf{Phylogenetic Graph:} This is a directed acyclic graph (DAG) rooted at an initial version. Each node encodes a complete snapshot of the codebase, while edges represent validated evolutionary steps. The graph induces a partial order over all code versions, enabling multiple developmental trajectories to coexist and recombine.

    \item \textbf{Git Backend:} The graph is implemented directly on top of Git, ensuring compatibility with standard tools such as GitHub. This design supports reproducibility, distributed development, version control, and fine-grained tracking of code changes.
\end{itemize}

This decentralized architecture enables fully asynchronous agent operation without centralized scheduling or inter-agent communication. Coordination emerges implicitly through shared interaction with the evolving version graph, where each agent’s contribution is grounded in the common structure rather than explicit messaging or shared memory.

\subsection{Code Evolution as Graph Expansion}

EvoGit conceptualizes software development as a continual expansion of a \textit{phylogenetic graph}. This structure facilitates safe, incremental, and traceable development by representing all version transitions explicitly:

\begin{itemize}
    \item Each node encodes a complete snapshot of the codebase at a specific version.
    \item A directed edge from $v_i$ to $v_j$ denotes that $v_j$ is a validated evolutionary successor of $v_i$ (i.e., $v_i \preceq v_j$).
    \item The \textit{maximal nodes} (i.e., those with no descendants) constitute the current evolutionary frontier and serve as candidate starting points for further development.
\end{itemize}

Agents drive the expansion of this graph by iteratively applying mutation and crossover operations, following an autonomous and localized decision loop:

\begin{enumerate}
    \item \textbf{Base Selection:} The agent selects a version from the current frontier to build upon.
    
    \item \textbf{Candidate Generation:}
    \begin{itemize}
        \item For \textbf{mutation}, the agent receives the design specification, current file structure, and a randomly chosen editable code segment. It then proposes a minor local change, e.g., modifying a function, inserting a conditional, or refactoring a few lines.
        \item For \textbf{crossover}, the agent selects two parent nodes (typically from divergent branches) and invokes an external three-way merge routine. A shared ancestor is identified, and an offspring version is synthesized via structural diffing and randomized conflict resolution.
    \end{itemize}
    
    \item \textbf{Candidate Evaluation:} The proposed version is locally compared against its parent(s) using diagnostics (e.g., compiler output, linter results, test reports) and semantic diffs. The agent determines whether the new version represents a meaningful improvement.
    
    \item \textbf{Graph Update:} If accepted, the candidate is added to the phylogenetic graph as a new node with appropriate parent edges. Otherwise, the proposal is discarded.
\end{enumerate}

This graph-expansion paradigm mirrors the fundamental mechanism of biological evolution, wherein complex organisms emerge not through sudden leaps, but through the gradual accumulation of small and viable variations. 
EvoGit follows the same principle: each code update is intentionally constrained to be both safe and beneficial, enabling robust, long-term development without centralized control:

\begin{itemize}
    \item \textbf{Incremental:} Each change is deliberately minor and localized, making it easier to validate and less likely to introduce unintended regressions.
    \item \textbf{Advantageous:} A new version must demonstrate clear improvement over its parent(s) to survive in the version lineage.
\end{itemize}

This \textit{small-step evolution} constraint is not merely a technical convenience, but it is a design principle rooted in evolutionary epistemology. It ensures that individual agents can make independent progress with minimal risk, while the system as a whole steadily converges toward increasingly sophisticated and coherent software artifacts. Over time, these micro-refinements accumulate into large-scale structural innovations, all transparently recorded within the evolving version graph.

\subsection{Coordination Through Structure}

Coordination in EvoGit emerges not from centralized control, explicit communication, or shared memory, but organically from the shared structure of the phylogenetic graph.
This mechanism parallels coordination in natural ecosystems, where independent organisms interact through shared environments and evolutionary pressures, rather than direct communication.

\begin{itemize}
    \item Agents operate asynchronously and independently, proposing code changes and contributing new versions to the shared phylogenetic graph.
    \item The partial order over code versions allows agents to infer relative quality: if $v_1 \preceq v_2$ and $v_2 \preceq v_3$, then by transitivity, $v_1 \preceq v_3$.
    \item Divergent branches (represented as incomparable nodes) preserve exploratory diversity and can later be recombined via crossover operations.
\end{itemize}

This structure-driven coordination resembles stigmergy in biological systems, where indirect interactions through a shared medium (e.g., pheromone trails in ants, environmental gradients in bacteria) enable complex collective behaviors. EvoGit’s version graph serves as both memory and medium, i.e., an evolving substrate that guides agent behavior without requiring synchronization or messaging.

A key design principle is that agent state is externalized: agents do not retain long-term memory. All progress and decisions are recorded in the version graph itself. This statelessness enhances robustness. If an agent fails (e.g., by producing invalid code or making poor comparisons), its failure is localized and harmless. As long as at least one agent remains functional, the overall system continues to evolve productively.

In this way, EvoGit mirrors real-world collaborative software development, where multiple developers work independently, contribute asynchronously, and occasionally merge their progress. The result is a resilient, transparent, and self-organizing development process, well-suited to large-scale, distributed, and partially unreliable environments.

\subsection{Sparse Human-in-the-Loop Oversight}

While EvoGit is primarily designed for autonomous operation, sparse human feedback plays a vital role in aligning development with high-level goals. Human involvement occurs in two key capacities:

\begin{itemize}
\item \textbf{Initialization:} A human provides the initial design intent, defines development objectives, and supplies any essential seed assets or files required to bootstrap the project.
\item \textbf{Strategic Feedback:} At periodic checkpoints, the human reviews the phylogenetic graph to prune unproductive branches, promote promising directions, or insert lightweight annotations that influence agent behavior. Such interventions help clarify ambiguous objectives and guide the trajectory of exploration.
\end{itemize}

This oversight model reflects real-world software workflows, in which product managers steer development through strategic guidance rather than micromanagement or direct code contributions. In EvoGit, the human functions as a high-level curator (i.e., intervening only when necessary) while autonomous agents execute the majority of iterative work. This balance between autonomy and oversight enables both creative freedom and directional coherence.

\section{Experiments}

\subsection{Tasks and Settings}

To assess the capabilities of {EvoGit} in realistic development contexts, we evaluate it on two representative software construction tasks. In both cases, the codebase is initialized with minimal scaffolding and evolved via asynchronous agent operations.

\paragraph{Task 1: Web Application Development}
This task evaluates EvoGit's ability to autonomously construct a \textit{functional promotional website for a research project}. The target application requires coordinated manipulation of HTML, CSS, and JavaScript components to achieve an interactive and aesthetically coherent user interface. Agents receive only the project brief and initial folder structure, but not predefined templates, code hints, or visual rendering.
A population of 16 autonomous coding agents operates in parallel, each performing up to 120 iterations. Every 10 iterations, sparse human feedback is introduced to guide direction, in the form of high-level structural suggestions, branch pruning, or promotion of specific versions. This setup evaluates EvoGit’s capacity for decentralized layout synthesis, interactive logic coordination, and semantic consistency without explicit supervision.

\paragraph{Task 2: Meta-Level Code Synthesis}
This task evaluates EvoGit's capacity to support recursive and system-level code generation by evolving an \textit{automated algorithm design pipeline}. Specifically, the target artifact is not a direct solver for the bin-packing optimization problem, but a meta-system that autonomously constructs such solvers using a LLM. In effect, EvoGit is tasked with synthesizing a program that itself serves as a meta-coding agent.
The resulting system must coordinate multiple components: LLM-based prompt construction, candidate code generation, fitness evaluation, and iterative refinement, all of which are integrated into a coherent and self-contained framework. 
This represents a higher-order development scenario, where the evolved software is not an application per se, but a general-purpose, self-adaptive problem-solving tool.
The experiment also runs 16 agents over 120 iterations, with human feedback provided every 20 iterations.

Complete implementation details, including the evaluation procedure, agent prompts, and crossover mechanisms, are provided in Appendix~\ref{appendix:procedure}.

\subsection{Experimental Results}

In both tasks, the initial version was minimally seeded by a human and all subsequent development was conducted by decentralized coding agents operating under the EvoGit framework.
Complete Git repositories are made publicly available for inspection and reproducibility.

Thanks to EvoGit’s Git-based version control, every modification (from the earliest stub to the final implementation) is permanently recorded in the phylogenetic graph. This enables transparent auditing, reproducibility, and replay of the entire development trajectory.
To facilitate detailed inspection and full reproducibility, we release the complete Git repositories for both tasks: Task 1\footnote{\url{https://github.com/BillHuang2001/evogit_web}} and Task 2\footnote{\url{https://github.com/BillHuang2001/evogit_llm}}. The corresponding phylogenetic graphs, which capture the complete version history, can be explored via GitHub’s network view for Task 1\footnote{\url{https://github.com/BillHuang2001/evogit_web/network}} and Task 2\footnote{\url{https://github.com/BillHuang2001/evogit_llm/network}}.

\textbf{Web Applicaiton Development:} 

\begin{figure}[htpb]
    \centering
    \begin{subfigure}{0.7\textwidth}
        \centering
        \includegraphics[trim={0 800 0 0},clip,width=\textwidth]{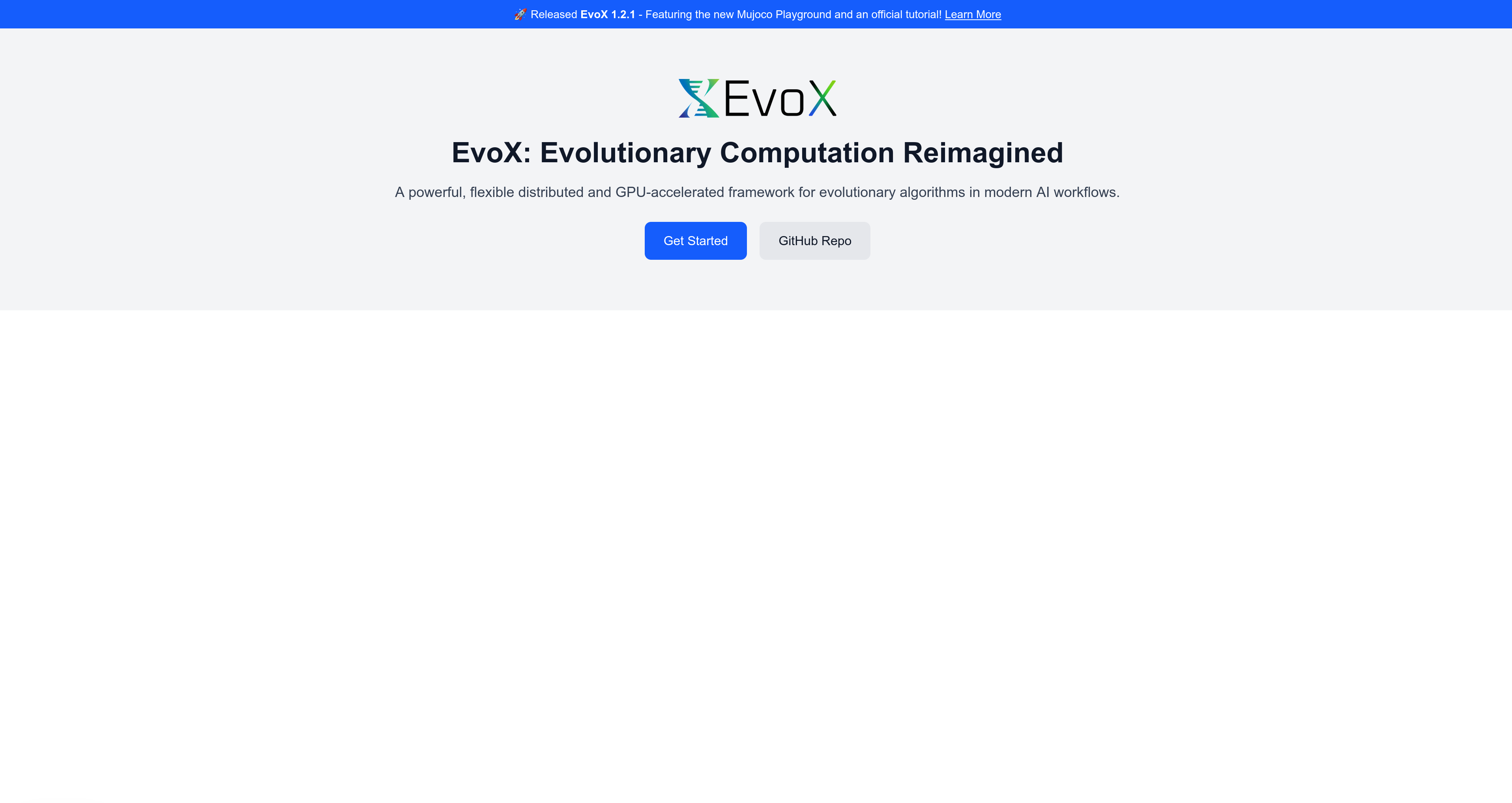}
        \caption{Early stage}
        \label{fig:evogit_web_early}
    \end{subfigure}
    
    \begin{subfigure}{0.7\textwidth}
        \centering
        \includegraphics[width=\textwidth]{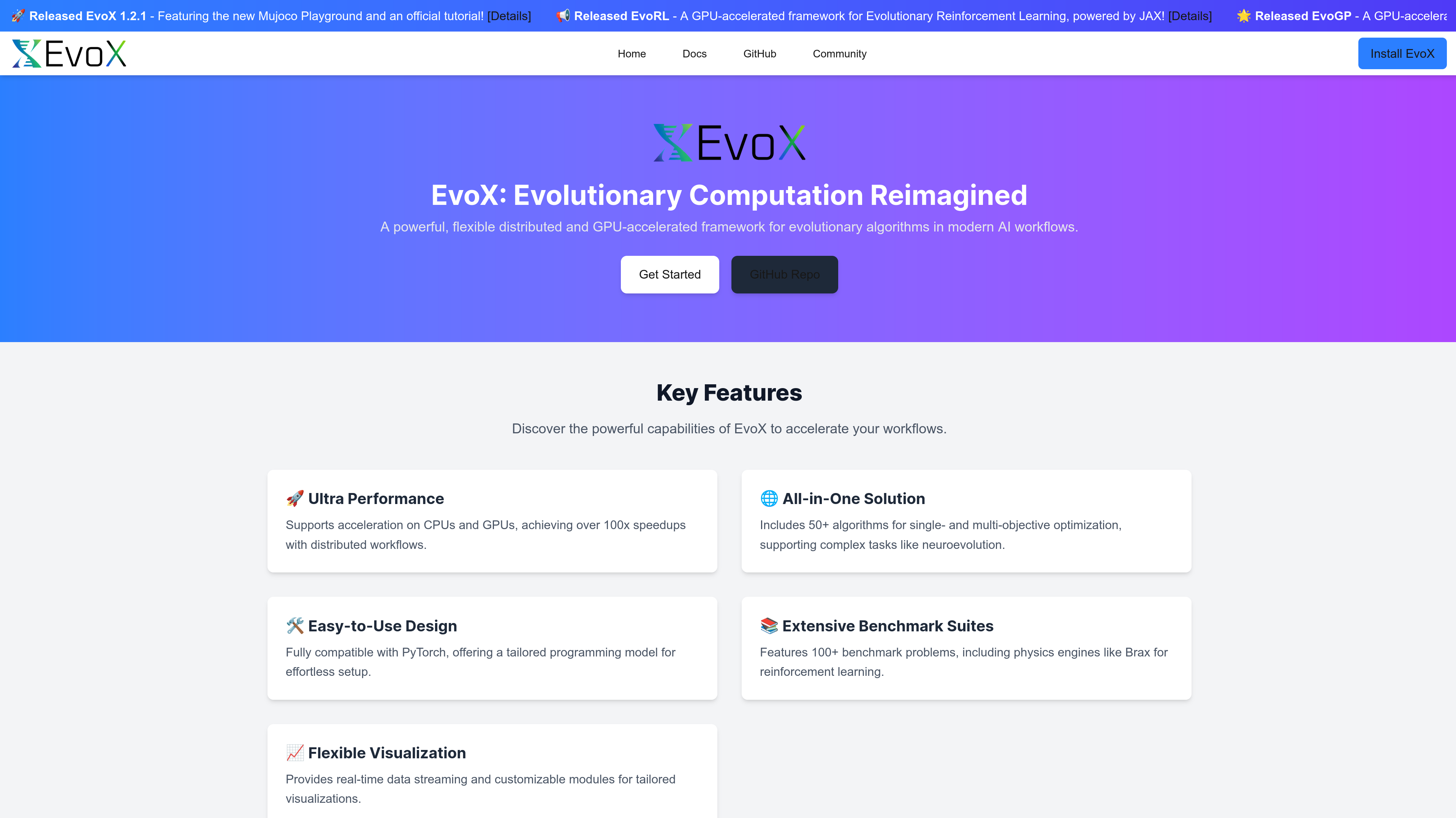}
        \caption{Middle stage}
        \label{fig:evogit_web_mid}
    \end{subfigure}
    
    \begin{subfigure}{0.7\textwidth}
        \centering
        \includegraphics[width=\textwidth]{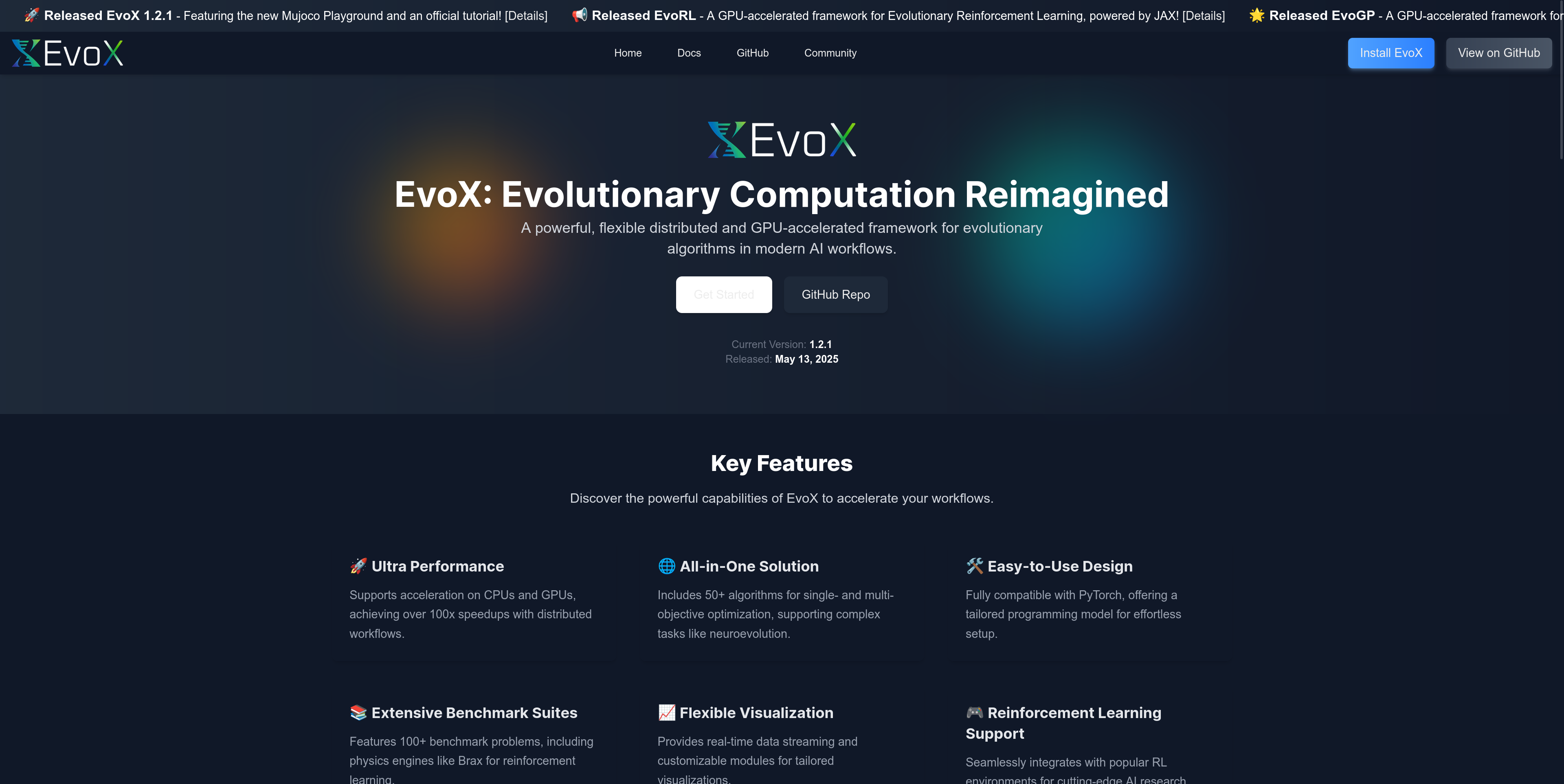}
        \caption{Final stage}
        \label{fig:evogit_web_final}
    \end{subfigure}
    
\caption{
Snapshots from the web development process by EvoGit (Task 1).  
(a) Early stage: broken layout with only a few scattered components.  
(b) Middle stage: structural elements begin to take shape, but styling remains minimal.  
(c) Final stage: fully functional and visually polished interface, with support for both light and dark modes.
}
    \label{fig:evogit_web}
\end{figure}

Figure~\ref{fig:evogit_web} illustrates the development trajectory of a web application created entirely by autonomous coding agents. 
The project was initialized by a human using the \texttt{Next.js} scaffolding tool, along with a set of provided image assets. 
All subsequent code was generated exclusively by agents, without any visual or image-based feedback, interacting purely through text.

The final product exhibits a clean and functional design that adapts seamlessly to both light and dark modes.
Core UI components are correctly implemented and styled, with smooth animations enhancing the user experience. 
This showcases EvoGit’s capability to operate effectively within modern web development frameworks.
Furthermore, the resulting codebase demonstrates good modular structure and separation of concerns, aligning with sound software engineering practices.

\begin{figure}[htpb]
\centering
\includegraphics[width=0.5\linewidth]{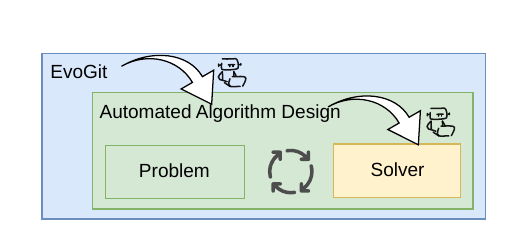}
\caption{Schematic of meta-level code synthesis (Task 2). EvoGit is used to develop a program that itself evolves a solver to the bin-packing problem using an LLM in the loop.}
\label{fig:task2_explain}
\end{figure}

\textbf{Meta-Level Code Synthesis:}

This task evaluates EvoGit’s ability to generate not just application-specific solutions, but an entire system for automated algorithm design. The goal is to evolve a program that orchestrates an outer-loop evolutionary process, which leverages a language model to iteratively synthesize and refine solvers for the bin-packing optimization problem. A schematic of this meta-level system is shown in Fig.~\ref{fig:task2_explain}.

The final product is a functional Python pipeline that coordinates prompt generation, LLM interaction, solution evaluation, and iterative refinement.
Remarkably, several standard engineering features (e.g., input validation, logging, and exception handling) emerged without explicit instruction, reflecting the agents’ capacity to generalize best practices.

While the generated project is narrower in scope than EvoGit itself, it demonstrates the feasibility of recursive development: using EvoGit to build tools that, in turn, perform automated code synthesis. 
This opens a path toward self-improving, general-purpose development agents.

\begin{lstlisting}[
    float=t,
  language=Python,
  caption={\texttt{best\_solution.py} found by the LLM-based code evolution program generated by the EvoGit framework.}
]
def bin_packing_solver(items: list[float], budget: int) -> list[int]:
    import time
    
    # Validate input
    if not isinstance(items, list) or not all(isinstance(i, (int, float)) and 0 <= i <= 1 for i in items):
        raise ValueError("Items must be a list of floats between 0 and 1.")
    if not isinstance(budget, int) or budget <= 0:
        raise ValueError("Budget must be a positive integer.")
    
    start_time = time.time()
    
    # Sort items in descending order for First-Fit Decreasing heuristic
    sorted_items = sorted(items, reverse=True)

    # Prepare bins to store the assigned items
    bins = []
    assignments = [-1] * len(sorted_items)

    for index, weight in enumerate(sorted_items):
        if time.time() - start_time > (budget / 1000) * 0.9:
            break  # Leave a buffer time
        
        placed = False
        for bin_index, bin in enumerate(bins):
            if sum(bin) + weight <= 1:
                bin.append(weight)
                assignments[index] = bin_index
                placed = True
                break
        
        if not placed:
            bins.append([weight])
            assignments[index] = len(bins) - 1

    # Store the best found solution
    best_solution = assignments.copy()
    initial_bins_count = len(bins)

    # Iterative refinement
    while time.time() - start_time <= (budget / 1000) * 0.9:
        current_bins = []
        current_assignments = [-1] * len(sorted_items)

        for index, weight in enumerate(sorted_items):
            if time.time() - start_time > (budget / 1000) * 0.9:
                break
            
            placed = False
            for bin_index, bin in enumerate(current_bins):
                if sum(bin) + weight <= 1:
                    bin.append(weight)
                    current_assignments[index] = bin_index
                    placed = True
                    break
            
            if not placed:
                current_bins.append([weight])
                current_assignments[index] = len(current_bins) - 1

        # If the new solution uses fewer bins, update the best solution
        if len(current_bins) < initial_bins_count:
            best_solution = current_assignments
            initial_bins_count = len(current_bins)
    
    return best_solution
    \end{lstlisting}
    \label{lst:task2}

\section{Conclusion}

EvoGit reimagines software development as an evolutionary process over a Git-based phylogenetic graph, enabling decentralized and asynchronous collaboration without scalar rewards or centralized control. By organizing all versions as nodes in a directed acyclic graph, EvoGit ensures traceability, scalability, and emergent coordination among agents.
Although this work focuses on autonomous code generation, the framework naturally generalizes to other domains that involve structured and versioned artifacts. The explicit lineage embedded in the graph allows agents to reason over historical trajectories, supporting more informed and context-aware evolution.
Promising directions for future work include adaptive pruning of the version graph, automated evaluation of divergent branches, and integration with more capable agents equipped with temporal reasoning and long-horizon planning.
Overall, EvoGit offers a new paradigm for collaborative system construction, which is grounded not in synchronized workflows, but in the open-ended exploration of an interpretable and evolving version space.

\FloatBarrier
\bibliography{iclr2025_conference}

\begin{thebibliography}{18}
\providecommand{\natexlab}[1]{#1}
\providecommand{\url}[1]{\texttt{#1}}
\expandafter\ifx\csname urlstyle\endcsname\relax
  \providecommand{\doi}[1]{doi: #1}\else
  \providecommand{\doi}{doi: \begingroup \urlstyle{rm}\Url}\fi

\bibitem[Hong et~al.(2024)Hong, Zhuge, Chen, Zheng, Cheng, Wang, Zhang, Wang, Yau, Lin, Zhou, Ran, Xiao, Wu, and Schmidhuber]{hong2024metagpt}
Sirui Hong, Mingchen Zhuge, Jonathan Chen, Xiawu Zheng, Yuheng Cheng, Jinlin Wang, Ceyao Zhang, Zili Wang, Steven Ka~Shing Yau, Zijuan Lin, Liyang Zhou, Chenyu Ran, Lingfeng Xiao, Chenglin Wu, and J{\"u}rgen Schmidhuber.
\newblock Meta{GPT}: Meta programming for a multi-agent collaborative framework.
\newblock In \emph{The Twelfth International Conference on Learning Representations}, 2024.
\newblock URL \url{https://openreview.net/forum?id=VtmBAGCN7o}.

\bibitem[Islam et~al.(2024)Islam, Ali, and Parvez]{islam2024mapcoder}
Md~Ashraful Islam, Mohammed~Eunus Ali, and Md~Rizwan Parvez.
\newblock Mapcoder: Multi-agent code generation for competitive problem solving.
\newblock \emph{arXiv preprint arXiv:2405.11403}, 2024.

\bibitem[Lehman et~al.(2023)Lehman, Gordon, Jain, Ndousse, Yeh, and Stanley]{lehman2023evolution}
Joel Lehman, Jonathan Gordon, Shawn Jain, Kamal Ndousse, Cathy Yeh, and Kenneth~O Stanley.
\newblock Evolution through large models.
\newblock In \emph{Handbook of Evolutionary Machine Learning}, pp.\  331--366. Springer, 2023.

\bibitem[Liu et~al.(2024)Liu, Tong, Yuan, Lin, Luo, Wang, Lu, and Zhang]{liu2023algorithm}
Fei Liu, Xialiang Tong, Mingxuan Yuan, Xin Lin, Fu~Luo, Zhenkun Wang, Zhichao Lu, and Qingfu Zhang.
\newblock Evolution of heuristics: Towards efficient automatic algorithm design using large language model.
\newblock In \emph{Proceedings of the 41st International Conference on Machine Learning}, pp.\  1--9, 2024.

\bibitem[Minaee et~al.(2024)Minaee, Mikolov, Nikzad, Chenaghlu, Socher, Amatriain, and Gao]{minaee_large_2024}
Shervin Minaee, Tomas Mikolov, Narjes Nikzad, Meysam Chenaghlu, Richard Socher, Xavier Amatriain, and Jianfeng Gao.
\newblock Large {Language} {Models}: {A} {Survey}, February 2024.
\newblock URL \url{http://arxiv.org/abs/2402.06196}.
\newblock arXiv:2402.06196 [cs].

\bibitem[OpenAI et~al.(2024)OpenAI, Achiam, Adler, Agarwal, Ahmad, Akkaya, Aleman, Almeida, Altenschmidt, Altman, Anadkat, Avila, et~al.]{openai_gpt-4_2024}
OpenAI, Josh Achiam, Steven Adler, Sandhini Agarwal, Lama Ahmad, Ilge Akkaya, Florencia~Leoni Aleman, Diogo Almeida, Janko Altenschmidt, Sam Altman, Shyamal Anadkat, Red Avila, et~al.
\newblock {GPT}-4 {Technical} {Report}, March 2024.
\newblock URL \url{http://arxiv.org/abs/2303.08774}.
\newblock arXiv:2303.08774 [cs].

\bibitem[Qian et~al.(2023)Qian, Liu, Liu, Chen, Dang, Li, Yang, Chen, Su, Cong, Xu, Li, Liu, and Sun]{chatdev_2023}
Chen Qian, Wei Liu, Hongzhang Liu, Nuo Chen, Yufan Dang, Jiahao Li, Cheng Yang, Weize Chen, Yusheng Su, Xin Cong, Juyuan Xu, Dahai Li, Zhiyuan Liu, and Maosong Sun.
\newblock Chatdev: Communicative agents for software development.
\newblock \emph{arXiv preprint arXiv:2307.07924}, 2023.
\newblock URL \url{https://arxiv.org/abs/2307.07924}.

\bibitem[Romera-Paredes et~al.(2023)Romera-Paredes, Barekatain, Novikov, Balog, Kumar, Dupont, Ruiz, Ellenberg, Wang, Fawzi, et~al.]{romera2023mathematical}
Bernardino Romera-Paredes, Mohammadamin Barekatain, Alexander Novikov, Matej Balog, M~Pawan Kumar, Emilien Dupont, Francisco~JR Ruiz, Jordan~S Ellenberg, Pengming Wang, Omar Fawzi, et~al.
\newblock Mathematical discoveries from program search with large language models.
\newblock \emph{Nature}, pp.\  1--3, 2023.

\bibitem[Stanley \& Miikkulainen(2002)Stanley and Miikkulainen]{stanley2002evolving}
Kenneth~O Stanley and Risto Miikkulainen.
\newblock Evolving neural networks through augmenting topologies.
\newblock \emph{Evolutionary computation}, 10\penalty0 (2):\penalty0 99--127, 2002.

\bibitem[Stanley et~al.(2019)Stanley, Clune, Lehman, and Miikkulainen]{stanley2019designing}
Kenneth~O. Stanley, Jeff Clune, Joel Lehman, and Risto Miikkulainen.
\newblock Designing neural networks through neuroevolution.
\newblock \emph{Nature Machine Intelligence}, 1\penalty0 (1):\penalty0 24--35, 2019.

\bibitem[Team et~al.(2023)Team, Anil, Borgeaud, Wu, Alayrac, Yu, Soricut, Schalkwyk, Dai, Hauth, Millican, Silver, Petrov, Johnson, Antonoglou, Schrittwieser, Glaese, Chen, et~al.]{gemini_team_gemini_2023}
Gemini Team, Rohan Anil, Sebastian Borgeaud, Yonghui Wu, Jean-Baptiste Alayrac, Jiahui Yu, Radu Soricut, Johan Schalkwyk, Andrew~M. Dai, Anja Hauth, Katie Millican, David Silver, Slav Petrov, Melvin Johnson, Ioannis Antonoglou, Julian Schrittwieser, Amelia Glaese, Jilin Chen, et~al.
\newblock Gemini: {A} {Family} of {Highly} {Capable} {Multimodal} {Models}, December 2023.
\newblock URL \url{http://arxiv.org/abs/2312.11805}.
\newblock arXiv:2312.11805 [cs].

\bibitem[Wang et~al.(2024)Wang, Ma, Feng, Zhang, Yang, Zhang, Chen, Tang, Chen, Lin, Zhao, Wei, and Wen]{wang_survey_2024}
Lei Wang, Chen Ma, Xueyang Feng, Zeyu Zhang, Hao Yang, Jingsen Zhang, Zhiyuan Chen, Jiakai Tang, Xu~Chen, Yankai Lin, Wayne~Xin Zhao, Zhewei Wei, and Ji-Rong Wen.
\newblock A {Survey} on {Large} {Language} {Model} based {Autonomous} {Agents}.
\newblock \emph{Frontiers of Computer Science}, 18\penalty0 (6):\penalty0 186345, December 2024.
\newblock ISSN 2095-2228, 2095-2236.
\newblock \doi{10.1007/s11704-024-40231-1}.
\newblock URL \url{http://arxiv.org/abs/2308.11432}.
\newblock arXiv:2308.11432 [cs].

\bibitem[Xi et~al.(2023)Xi, Chen, Guo, He, Ding, Hong, Zhang, Wang, Jin, Zhou, Zheng, Fan, Wang, Xiong, Zhou, Wang, Jiang, Zou, Liu, Yin, Dou, Weng, Cheng, Zhang, Qin, Zheng, Qiu, Huang, and Gui]{xi2023rise}
Zhiheng Xi, Wenxiang Chen, Xin Guo, Wei He, Yiwen Ding, Boyang Hong, Ming Zhang, Junzhe Wang, Senjie Jin, Enyu Zhou, Rui Zheng, Xiaoran Fan, Xiao Wang, Limao Xiong, Yuhao Zhou, Weiran Wang, Changhao Jiang, Yicheng Zou, Xiangyang Liu, Zhangyue Yin, Shihan Dou, Rongxiang Weng, Wensen Cheng, Qi~Zhang, Wenjuan Qin, Yongyan Zheng, Xipeng Qiu, Xuanjing Huang, and Tao Gui.
\newblock The rise and potential of large language model based agents: A survey, 2023.

\bibitem[Yang et~al.(2024{\natexlab{a}})Yang, Yang, Hui, Zheng, Yu, Zhou, Li, Li, Liu, Huang, Dong, Wei, Lin, Tang, Wang, Yang, Tu, Zhang, Ma, Yang, Xu, Zhou, Bai, He, Lin, Dang, Lu, Chen, Yang, Li, Xue, Ni, Zhang, Wang, Peng, Men, Gao, Lin, Wang, Bai, Tan, Zhu, Li, Liu, Ge, Deng, Zhou, Ren, Zhang, Wei, Ren, Liu, Fan, Yao, Zhang, Wan, Chu, Liu, Cui, Zhang, Guo, and Fan]{yang_qwen2_2024}
An~Yang, Baosong Yang, Binyuan Hui, Bo~Zheng, Bowen Yu, Chang Zhou, Chengpeng Li, Chengyuan Li, Dayiheng Liu, Fei Huang, Guanting Dong, Haoran Wei, Huan Lin, Jialong Tang, Jialin Wang, Jian Yang, Jianhong Tu, Jianwei Zhang, Jianxin Ma, Jianxin Yang, Jin Xu, Jingren Zhou, Jinze Bai, Jinzheng He, Junyang Lin, Kai Dang, Keming Lu, Keqin Chen, Kexin Yang, Mei Li, Mingfeng Xue, Na~Ni, Pei Zhang, Peng Wang, Ru~Peng, Rui Men, Ruize Gao, Runji Lin, Shijie Wang, Shuai Bai, Sinan Tan, Tianhang Zhu, Tianhao Li, Tianyu Liu, Wenbin Ge, Xiaodong Deng, Xiaohuan Zhou, Xingzhang Ren, Xinyu Zhang, Xipin Wei, Xuancheng Ren, Xuejing Liu, Yang Fan, Yang Yao, Yichang Zhang, Yu~Wan, Yunfei Chu, Yuqiong Liu, Zeyu Cui, Zhenru Zhang, Zhifang Guo, and Zhihao Fan.
\newblock Qwen2 {Technical} {Report}, July 2024{\natexlab{a}}.
\newblock URL \url{https://arxiv.org/abs/2407.10671v4}.

\bibitem[Yang et~al.(2024{\natexlab{b}})Yang, Jimenez, Wettig, Lieret, Yao, Narasimhan, and Press]{yang_swe-agent_2024}
John Yang, Carlos~E. Jimenez, Alexander Wettig, Kilian Lieret, Shunyu Yao, Karthik Narasimhan, and Ofir Press.
\newblock {SWE}-agent: {Agent}-{Computer} {Interfaces} {Enable} {Automated} {Software} {Engineering}, November 2024{\natexlab{b}}.
\newblock URL \url{http://arxiv.org/abs/2405.15793}.
\newblock arXiv:2405.15793 [cs].

\bibitem[Yao et~al.(2023)Yao, Zhao, Yu, Du, Shafran, Narasimhan, and Cao]{yao2023react}
Shunyu Yao, Jeffrey Zhao, Dian Yu, Nan Du, Izhak Shafran, Karthik Narasimhan, and Yuan Cao.
\newblock {ReAct}: Synergizing reasoning and acting in language models.
\newblock In \emph{International Conference on Learning Representations (ICLR)}, 2023.

\bibitem[Ye et~al.(2024)Ye, Wang, Cao, and Song]{ye2024reevo}
Haoran Ye, Jiarui Wang, Zhiguang Cao, and Guojie Song.
\newblock Reevo: Large language models as hyper-heuristics with reflective evolution.
\newblock \emph{arXiv preprint arXiv:2402.01145}, 2024.

\bibitem[Zhao et~al.(2023)Zhao, Zhou, Li, Tang, Wang, Hou, Min, Zhang, Zhang, Dong, Du, Yang, Chen, Chen, Jiang, Ren, Li, Tang, Liu, Liu, Nie, and Wen]{zhao_survey_2023}
Wayne~Xin Zhao, Kun Zhou, Junyi Li, Tianyi Tang, Xiaolei Wang, Yupeng Hou, Yingqian Min, Beichen Zhang, Junjie Zhang, Zican Dong, Yifan Du, Chen Yang, Yushuo Chen, Zhipeng Chen, Jinhao Jiang, Ruiyang Ren, Yifan Li, Xinyu Tang, Zikang Liu, Peiyu Liu, Jian-Yun Nie, and Ji-Rong Wen.
\newblock A {Survey} of {Large} {Language} {Models}, November 2023.
\newblock URL \url{http://arxiv.org/abs/2303.18223}.
\newblock arXiv:2303.18223 [cs].

\end{thebibliography}
\bibliographystyle{iclr2025_conference}

\appendix
\appendix
\onecolumn
\section{Related Work}

\subsection{Coding Agents}

Recent advances have demonstrated the potential of LLM-powered agents to automate complex software engineering tasks through structured collaboration.
{ChatDev}~\cite{chatdev_2023} simulates a virtual software company by orchestrating role-based agents aligned with the waterfall model, covering phases such as design, coding, testing, and documentation. It highlights collective intelligence through scripted inter-agent communication.
{MetaGPT}~\cite{hong2024metagpt} formalizes such workflows via a meta-programming framework that encodes agent roles and produces modular, maintainable codebases through structured collaboration.
{SWE-agent}~\cite{yang_swe-agent_2024} introduces an Agent-Computer Interface (ACI) that enables agents to perform software engineering tasks directly on real systems, achieving strong results on standard benchmarks through low-level, actionable feedback.
{MapCoder}~\cite{islam2024mapcoder} focuses on competitive programming by employing a sequential pipeline of specialized agents for problem decomposition, planning, coding, and debugging.

While effective, these systems generally rely on synchronous orchestration or message-passing protocols, which limit scalability. In contrast, {EvoGit} enables fully asynchronous collaboration via a shared phylogenetic graph, avoiding centralized coordination and supporting emergent behavior through implicit version-based interaction.

\subsection{Code Evolution}

Framing code generation as an evolutionary process has also gained attention, where programs evolve through mutation, recombination, and selection.
Early foundations by Lehman et al.~\cite{lehman2023evolution} introduced open-endedness and novelty search in code evolution, emphasizing exploration over direct optimization.
This inspired subsequent approaches that adapt evolutionary computation for program synthesis.
{FunSearch}~\cite{romera2023mathematical} used LLM-generated functional programs as evolutionary units to discover novel mathematical results guided by correctness objectives.
{EoH}~\cite{liu2023algorithm} evolved heuristic components as algorithmic primitives, enabling automated algorithm design and tuning.
{ReEvo}~\cite{ye2024reevo} incorporated reflective self-evaluation and iterative refinement to promote code reuse and long-term improvement.

However, these methods often depend on scalar fitness metrics and evolve monolithic programs, limiting their applicability to collaborative, real-world software development. EvoGit addresses these gaps by enabling modular co-evolution and relying on structural ordering rather than scalar rewards.

\subsection{Lineage and Structural Memory}

A key insight from evolutionary computation is the value of temporal structure in guiding recombination. {NEAT}~\cite{stanley2002evolving,stanley2019designing} introduced historical markings to align genes during crossover, preserving innovation history in evolving neural networks.
EvoGit generalizes this concept to software development. Rather than aligning neurons, we use Git commits as history markers over source code, embedding structural memory in a directed acyclic version graph. This enables decentralized agents to evolve, recombine, and trace code contributions over time, which scales lineage tracking from individual solutions to collaborative, multi-agent development.

\section{Detailed Methodology}

This section introduces the foundational components of {EvoGit}, beginning with the formal definition of its core data structure—a Git-native phylogenetic graph grounded in partial order theory. 
We then describe the evolutionary operations that enable agents to propose, mutate, and recombine code versions asynchronously. 
Together, these elements support scalable, decentralized collaboration across agents and human developers.

\subsection{Data Structure}

\begin{figure}[htpb]
    \centering
    \includegraphics[width=\linewidth]{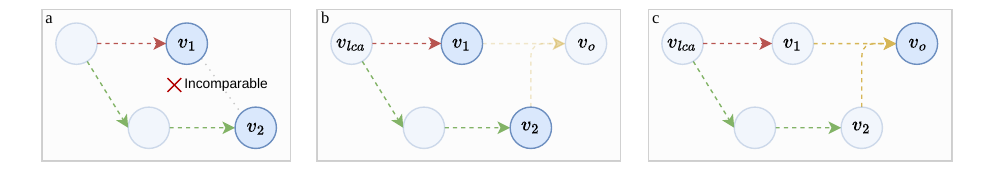}
    \caption{Partial order $\preceq$ over code versions. Each node denotes a version, and edges represent ancestry.  
    (a) Branching: $v_1$ and $v_2$ are incomparable, thus both are considered current best versions.  
    (b) Three-way merge: $v_o$ is created by recombining $v_1$ and $v_2$ with respect to their lowest common ancestor.  
    (c) If $v_1 \preceq v_o$ and $v_2 \preceq v_o$, then $v_o$ is a valid joint descendant and dominates both parents.}
    \label{fig:partial-order}
\end{figure}

At the core of EvoGit lies a \emph{phylogenetic graph}, i.e., a directed acyclic graph (DAG) that encodes the ancestry of all code versions.
Each node represents a complete snapshot of the codebase, while directed edges indicate derivation via mutation or merge. 
This structure naturally defines a partial order $\preceq$ over versions, where $v_i \preceq v_j$ implies that $v_i$ is an ancestor of $v_j$.

Figure~\ref{fig:partial-order} illustrates three key properties of this partial order. 
In (a), branching leads to multiple mutually incomparable versions, each representing a local optimum. 
In (b), a new version $v_o$ is synthesized through a three-way merge of $v_1$ and $v_2$, anchored by their nearest common ancestor. 
In (c), if both $v_1$ and $v_2$ precede $v_o$ under $\preceq$, the newly created version subsumes its parents and may be adopted as the new best candidate.

By adopting this formalism, EvoGit replaces scalar-valued fitness with structural dominance in the version space. 
This allows agents to collaborate indirectly through graph navigation and construction, without requiring explicit coordination or global objectives.

\subsubsection{Partially Ordered Code Versions}

Unlike domains with well-defined scalar objectives, real-world software development lacks a universal metric for ranking code versions. EvoGit addresses this by imposing a structural relation over versions based on their ancestry in the version control history.

Let $V$ denote the set of all code versions. We define a binary relation $\preceq$ over $V$ such that $a \preceq b$ if and only if $a$ is an ancestor of $b$ (either directly or via a transitive sequence of derivations). This induces a \textbf{partial order} over $V$ with the following properties:

\begin{itemize}
    \item \textbf{Reflexivity}: $a \preceq a$ for all $a \in V$.
    \item \textbf{Antisymmetry}: $a \preceq b$ and $b \preceq a$ imply $a = b$.
    \item \textbf{Transitivity}: $a \preceq b$ and $b \preceq c$ imply $a \preceq c$.
\end{itemize}

This partial order encodes lineage information, enabling agents to reason about dominance and inheritance without requiring scalar fitness values. Crucially, not all pairs are comparable: if $a$ and $b$ do not share an ancestor-descendant relationship, they are \textit{incomparable}. Such cases emerge naturally from concurrent development branches and represent potential candidates for evolutionary recombination. This structure reflects the open-ended, multi-directional nature of software evolution, where multiple paths may simultaneously explore and improve different aspects of the system.

Building on this relational foundation, we construct a global data structure (the phylogenetic graph) that records and organizes all versioning activity within EvoGit.

\subsubsection{Phylogenetic Graph}
\label{phylogentic-graph}

In EvoGit, the evolutionary history of software is represented as a \textit{phylogenetic graph}, i.e., a rooted directed acyclic graph (DAG) over the set of code versions $V$, which is induced by the partial order $\preceq$ formulated in the preceding subsection.

The graph is anchored at a unique root version $v_0$, representing the initial codebase. 
Each node corresponds to a code version, and a directed edge from $v_i$ to $v_j$ indicates that $v_j$ was derived from $v_i$ via modification, mutation, or recombination. 
For every version $v \in V$, there exists at least one directed path from $v_0$ to $v$, capturing its full lineage.

Unlike linear or tree-based version control histories, this graph supports concurrent, branching development paths. 
It enables multiple agents to explore diverse directions in parallel, while preserving the ancestral structure for traceability and recombination.
Crucially, the phylogenetic graph is grounded purely in structural ancestry and does not impose scalar fitness values. 
This avoids artificial total ordering and reflects the reality of software development, where comparative quality is often multi-dimensional and context-dependent.

One of the key advantages of this graph-based representation is that it provides a principled way to identify the most promising or actively evolving versions at any moment. 
This leads to the concept of \textit{best code versions}, which can be naturally formalized using maximal elements in the partial order induced by the graph.

\subsubsection{Best Code Versions}

Within the phylogenetic graph, the ``best'' code versions at a given point are characterized as \textit{maximal elements}, i.e., those that have no descendants under the partial order $\preceq$.

Formally, let $V$ denote the set of all code versions. A version $v \in V$ is a maximal element if there exists no $v' \in V$ such that $v \prec v'$, i.e., $v \preceq v'$ and $v \ne v'$:

\[
\text{Maximal}(V) = \{ v \in V \mid \nexists v' \in V, \; v \prec v' \}.
\]

Maximal elements represent the current evolutionary frontier, which are code versions that are not dominated by any successors. Since $\preceq$ is only a partial order, multiple such versions may coexist, reflecting diverse, concurrently evolving development branches.

\subsection{Operations}

This section introduces the core operations that act upon the phylogenetic graph and collectively drive EvoGit's evolutionary development process. These operations define how new code versions are generated, how divergent branches are reconciled, and how the system maintains and advances its evolutionary frontier.
The four fundamental operations (i.e., \textit{initialization}, \textit{mutation}, \textit{crossover}, and \textit{human intervention}) are detailed in the following subsections.

\begin{figure}[htpb]
    \centering
    \includegraphics[width=0.9\linewidth]{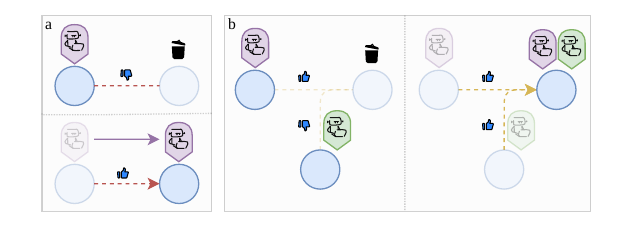}
    \caption{Illustration of the mutation and three-way crossover operations in EvoGit. 
    (a) Mutation: an agent selects a code version and proposes a change. If the change degrades functionality, the version is discarded and the agent continues from the original node. If the change is beneficial, the new version is added as a descendant, and the agent proceeds from the updated node.
    (b) Three-way crossover: two agents working on separate branches identify a common ancestor and attempt to merge their respective changes. If the resulting version is inferior to either parent, it is discarded. Otherwise, it is added to the graph as a valid descendant of both parents.}
    \label{fig:operations}
\end{figure}

Figure~\ref{fig:operations} provides a summary of EvoGit's two core mechanisms, mutation and crossover, which allow agents to continually push the evolutionary frontier while preserving the semantic integrity of the version graph. These operations form the backbone of open-ended, multi-agent code evolution within the EvoGit framework.

\subsubsection{Initialization}

The initialization operation establishes the starting point of the EvoGit evolutionary process. It creates a single root version, denoted $v_0$, which serves as the globally recognized origin of the phylogenetic graph. This root can be an \emph{empty scaffold}, a \emph{minimal working prototype}, or a \emph{curated template} provided by a human developer.

Typically, a human product manager oversees this phase, which includes configuring the development environment, loading essential build tools (e.g., compilers, linters), and defining initial project requirements. The initialized root provides a common foundation upon which all subsequent evolutionary modifications are built.

\subsubsection{Code Mutation}

Mutation is the primary mechanism through which individual agents explore the local neighborhood of a given code version. 
Starting from a current version $v_{\text{current}}$, an agent proposes a modified version $v_{\text{mut}}$ by applying a small and localized change. 
The neighborhood is defined by constraints such as:
\begin{itemize}
    \item Editing a specific section within an existing file.
    \item Creating a new file and inserting a defined code snippet.
\end{itemize}
These constraints ensure that mutations are incremental and non-disruptive, avoiding large-scale rewrites.

Once a mutation is applied, the agent performs a pairwise evaluation to determine whether $v_{\text{mut}}$ is an improvement over $v_{\text{current}}$. 
This decision relies on meta-information including build status, linter diagnostics, type-checking, and test results, which are evaluated both before and after the mutation. 
If $v_{\text{mut}}$ passes these checks and introduces no regressions, it is accepted into the phylogenetic graph as a descendant: $v_{\text{current}} \preceq v_{\text{mut}}$. 
The agent then continues evolving from $v_{\text{mut}}$. Otherwise, the mutation is discarded, and the agent remains at $v_{\text{current}}$.

Since mutations are localized, they fall well within the reasoning capabilities of modern LLMs, enabling reliable evaluations. 
When applied in parallel across agents, mutation supports broad and efficient exploration of the solution space, continuously extending the phylogenetic graph via diverse and independently evolved branches.

\subsubsection{Three-Way Crossover}
\label{context-aware-crossover}

Three-way crossover in EvoGit synthesizes a new code version by combining the divergent changes from two parent versions, guided by their shared evolutionary history. Unlike traditional crossover methods that naively splice code, EvoGit leverages the \textit{lowest common ancestor} (LCA) in the phylogenetic graph to ground the merge in structural and semantic context.

The LCA is defined as the deepest node in the graph from which both parents are reachable via directed paths. It represents the most recent common state of the codebase before the two parent versions diverged. By using the LCA as a reference point, EvoGit can interpret the specific modifications introduced along each evolutionary branch.

The three-way crossover proceeds as follows:
\begin{enumerate}
    \item \textbf{Identify Differences:} Compute the sets of changes introduced by each parent with respect to the LCA:
    \[
    \Delta_1 = \text{diff}(v_{\mathrm{lca}}, v_1), \quad
    \Delta_2 = \text{diff}(v_{\mathrm{lca}}, v_2)
    \].

    \item \textbf{Merge Changes:} Apply both $\Delta_1$ and $\Delta_2$ to the LCA to produce a candidate offspring $v_o$. If the changes are disjoint, they are merged directly. If they conflict (i.e., both parents modify the same region), conflicts are resolved via a randomized policy to promote diversity.

    \item \textbf{Validate Offspring:} The merged version is evaluated for structural and semantic correctness. If valid, and if it preserves ancestry from both parents (i.e., $v_1 \preceq v_o$ and $v_2 \preceq v_o$), the offspring is accepted and added to the phylogenetic graph.
\end{enumerate}
This crossover strategy ensures that the offspring captures the distinct contributions of both parents while maintaining coherent lineage. It requires no scalar fitness scores, but instead replies on structural integrity and ancestral consistency. Moreover, randomized conflict resolution introduces variation analogous to genetic recombination in biological evolution, enabling agents to explore novel combinations and directions in the solution space.

\subsubsection{Human Intervention}

As independent branches in the phylogenetic graph evolve over time, structural and semantic divergence among code versions may increase, reducing the likelihood of successful crossover. 
When divergence becomes excessive, agents may generate incompatible or unproductive versions, leading to wasted computational effort and fragmented development trajectories.

To prevent such fragmentation and preserve alignment with overarching project goals, EvoGit supports periodic \textit{human intervention}. 
A human participant (typically the product manager) can review the current state of the phylogenetic graph and selectively retain a subset of the maximal elements (i.e., $\text{Maximal}(V)$) as preferred directions for continued development, discarding the rest.

This pruning operation serves as a top-down corrective signal, refocusing agent exploration toward the chosen subset and restoring coherence across the population. It is particularly useful when automated operations such as mutation and crossover stagnate or produce inconsistent outputs.

Although EvoGit emphasizes autonomous multi-agent development, strategic human intervention provides a safeguard against long-term drift. This mechanism mirrors the real-world role of a product manager who periodically steers development toward high-level objectives, ensuring the system remains on track without micromanaging individual steps.

\subsection{Collaborative Development}

\begin{figure}[htpb]
    \centering
    \includegraphics[width=0.9\linewidth]{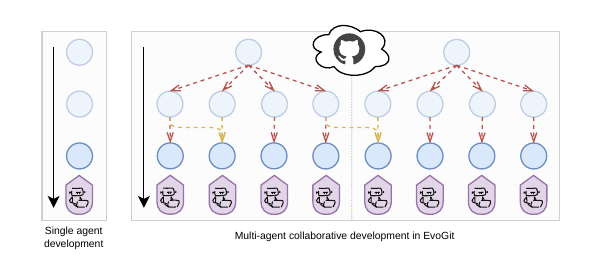}
    \caption{Distributed code evolution in EvoGit. Traditional workflows rely on linear, sequential development. In contrast, EvoGit enables multiple agents (potentially on different machines) to independently evolve code in parallel. Coordination emerges implicitly through the shared version graph, synchronized via Git-based platforms such as GitHub.}
    \label{fig:evogit-distributed}
\end{figure}

As shown in Fig.~\ref{fig:evogit-distributed}, traditional development workflows typically follow a linear trajectory, driven by a single agent guided by human instructions or scalar objectives.
EvoGit replaces this paradigm with decentralized, asynchronous collaboration. Multiple agents can work independently on different branches of the codebase, each exploring and extending a local region of the phylogenetic graph without requiring direct coordination, shared memory, or centralized scheduling.

This decentralized interaction is enabled by the phylogenetic graph structure introduced in Section~\ref{phylogentic-graph}, which acts as a globally shared substrate for recording and integrating all code versions. 
When agents wish to combine contributions, they can traverse the graph to identify a lowest common ancestor and invoke the three-way crossover operation described in Section~\ref{context-aware-crossover}.

Importantly, crossovers in EvoGit are not limited to tip versions. Any pair of valid nodes in the graph may be recombined, allowing agents to reuse historical contributions and re-integrate diverging development paths. This flexibility supports robust collaboration, even under conditions of long-term drift, and encourages the emergence of complex, reusable code structures through decentralized evolution.

\section{Implementation Detail}

This section outlines how EvoGit can be implemented by leveraging existing version control systems, particularly Git. Due to the conceptual alignment between evolutionary development and version control, Git serves as a natural substrate for EvoGit’s realization.
While EvoGit is not inherently dependent on Git, the framework’s core abstractions (e.g., code versions, lineages, and branching) are well-aligned with Git’s data model and operational semantics. Thus, Git provides a practical and intuitive backend for implementation.\footnote{Although Git offers a convenient foundation, EvoGit is agnostic to specific version control tools. Its framework can be realized independently of Git, provided equivalent functionality is supported. Moreover, Git’s default line-based diff and merge mechanisms are not central to EvoGit’s design; EvoGit can incorporate alternative differencing algorithms and custom merge drivers as needed.}
We begin by revisiting the mapping between EvoGit’s conceptual structures and Git primitives, and then describe how Git’s native operations are employed to support mutation, crossover, and distributed collaboration.

\subsection{Data Structure Mapping}

EvoGit’s core data structures, particularly the phylogenetic graph, are directly mapped onto Git primitives, enabling seamless integration with existing development workflows and tooling. This mapping supports interoperability and lowers the barrier to adoption for both human developers and autonomous agents.

\begin{itemize}
    \item \textbf{Code Version} corresponds to a Git \texttt{Commit}, created using the standard \texttt{git commit} operation. Each commit is uniquely identified by a SHA-1 or SHA-256 hash, which EvoGit encodes into a binary vector to index the version within the evolutionary graph.

    \item \textbf{Individual} is represented by a Git \texttt{Reference}\footnote{Often abbreviated as \texttt{ref} in Git terminology}, which points to a specific commit. In EvoGit, each agent operates on its own \texttt{Branch}, which serves as a persistent identifier for that agent’s evolutionary lineage.

    \item \textbf{Population} is modeled as the set of active branches. At any time, the current population of individuals can be inspected using commands like \texttt{git branch} or through custom queries.

    \item \textbf{Feedback and Metadata} are stored using Git \texttt{Notes}. This mechanism attaches auxiliary information to commits (e.g., build logs, test outcomes, and human annotations) without altering the commit content itself.
\end{itemize}

By grounding its abstractions in Git, EvoGit ensures compatibility with mainstream development practices while extending them to support large-scale, decentralized evolutionary computation over codebases.

\subsection{Build Tool Integration}

EvoGit integrates with standard build tools to gather feedback and guide evolutionary decisions in realistic development environments. In our initial setup, we demonstrate EvoGit on a \texttt{Next.js}\footnote{\url{https://nextjs.org/}} project, a modern web framework based on React. Feedback is automatically extracted from \texttt{ESLint}, which performs syntax checking, code style enforcement, and static analysis. The diagnostics are structured into metadata and attached to the corresponding commit as Git notes.
This mechanism allows agents to make informed decisions without relying on manually crafted scalar reward functions. Each code mutation or crossover is evaluated against practical criteria (e.g., build success, code quality, and type correctness) ensuring that only valid and meaningful variants propagate in the evolutionary process.

We further demonstrate EvoGit on a Python project using the \texttt{uv}\footnote{\url{https://docs.astral.sh/uv/}} build system and the \texttt{pyright}\footnote{\url{https://github.com/microsoft/pyright}} static type checker. \texttt{uv} provides isolated and reproducible environments, while \texttt{ruff} offers fast, rule-based linting. As with JavaScript, tool feedback is captured and attached to each commit, enabling language-agnostic, traceable evolution.

These settings highlight EvoGit’s flexibility in integrating with heterogeneous ecosystems and utilizing domain-specific tools to ensure code quality and functional correctness during autonomous evolution.

\subsection{Pairwise Comparison}

Pairwise comparison is the core mechanism in EvoGit for assessing whether a newly proposed code version improves upon its predecessor. Evaluations are conducted in two complementary modes:
\begin{itemize}
    \item \textbf{Direct Comparison:} Each new version is compared against its immediate parent(s) by analyzing the code diff and aggregating diagnostics from build tools such as compilers and linters. If available, lightweight human feedback is also incorporated. These signals are passed to an LLM-based judge, which returns a binary decision on whether the new version constitutes an improvement.

    \item \textbf{Transitive Comparison:} Since EvoGit’s version history is DAG, it enables transitive reasoning: if $v_1$ is better than $v_0$, and $v_2$ is better than $v_1$, then $v_2$ is inferred to be better than $v_0$. Self-comparisons are assumed to be neutral ($v \succeq v$). This inference reduces redundant evaluations and allows agents to track progress over long evolutionary paths without re-examining every intermediate step.
\end{itemize}

Since EvoGit restricts mutations to be small and localized, judgments remain simple and reliable. Minor edits are unlikely to introduce complex trade-offs, making binary decisions more robust. Assuming correctness of direct comparisons, all transitive inferences are logically sound.
Together, these mechanisms provide an efficient and principled evaluation framework, supporting autonomous and incremental code evolution without relying on scalar fitness functions or continuous human supervision.

\section{Experimental Setup}
\label{appendix:procedure}

As EvoGit is designed to tackle real-world, open-ended software development tasks, the experimental setup does not follow any standardized benchmark protocol or rely on predefined numerical metrics.
Instead, to ensure reproducibility, minimize human bias, and rigorously evaluate the framework’s effectiveness, we establish a strict and transparent protocol governing both human involvement and agent behavior.

\subsection{Human-Agent Interaction Protocol}

\textbf{Human Protocol:}  
A single human participant assumes the role of a product manager, offering high-level design intent and feedback. 
To prevent overfitting and manual intervention, the following rules are enforced:
\begin{itemize}
    \item Design instructions must remain abstract and high-level, referencing only necessary APIs or environmental constraints (e.g., available packages), without suggesting implementation details.
    \item The human is allowed to initialize the project (e.g., scaffold the codebase) but must not write or modify code thereafter.
    \item Interaction is permitted only at fixed intervals (e.g., every $n$ iterations), during which the human selects one preferred version from the current frontier and optionally provides brief qualitative feedback.
    \item Feedback is restricted to observable behaviors such as functionality (e.g., runtime errors) or presentation (e.g., layout flaws), and may not include code-level advice.
\end{itemize}

\textbf{Agent Protocol:}  
Agents operate independently and without memory. Each mutation step is executed with minimal context, consisting of:
\begin{itemize}
    \item The original task description provided by the human.
    \item The directory and file structure of the current version.
    \item A randomly selected editable region (up to 128 lines) from a randomly chosen file in the codebase.
\end{itemize}

Two tasks are conducted under this controlled setup. In both cases, we deploy 16 agents in parallel for 120 iterations. The main difference lies in the frequency of human feedback:
\begin{itemize}
    \item {Web Application Development (Task 1):} Human feedback is provided every 10 iterations to reflect the greater subjectivity of web tasks (e.g., visual design, user experience).
    \item {Meta-Level Code Synthesis (Task 2):} Feedback is provided every 20 iterations, as this task focuses on functional correctness and measurable outcomes, which require less frequent human input.
\end{itemize}

This experimental setup ensures that the vast majority of code contributions arise autonomously from the multi-agent evolutionary process, guided only by sparse human preferences. By strictly limiting both the scope of agent context and the frequency of human input, the procedure isolates EvoGit’s capacity for open-ended, collaborative code generation without any human-written code beyond the initial prompt.

\subsection{Agent Allocation}

A key property of EvoGit is that the number of actively developing agents always upper-bounds the number of current frontier solutions, denoted as $|\text{Maximal}(V)|$.
Starting from a single root version, new code versions are produced via mutation (localized changes) or crossover (merging lineages). Each new version is only accepted if it is no worse than its parent(s), preserving a partial order over the version space.
This design ensures that every frontier node (i.e., maximal element in the phylogenetic graph) has been discovered and maintained by at least one agent. 
As a result, the frontier’s growth is intrinsically tied to agent activity, preventing uncontrolled divergence and ensuring efficient allocation of exploration efforts.

\section{Discussion on Reproducibility}

While we impose strict constraints on human involvement and agent behavior, complete determinism in the EvoGit framework is inherently unattainable due to several sources of stochasticity:
\begin{itemize}
\item The output of LLMs is inherently non-deterministic, even with fixed prompts and seeds.
\item During crossover, conflicts between code segments are resolved through randomized heuristics.
\item Both the target file and the editable region selected for mutation are chosen randomly.
\end{itemize}

Importantly, this controlled randomness is not a flaw but a necessary feature of the framework. 
Deterministic mutation would lead to identical offspring for a given context, drastically limiting diversity and diminishing the exploratory capacity of the evolutionary process. 
By embracing stochasticity (particularly in LLM outputs), we preserve evolutionary variability, which is essential for navigating the vast and uncertain search space of open-ended code generation.

\section{Discussion on the Phylogenetic Graph in EvoGit vs. the Standard Git Commit Graph}

Although EvoGit is built atop Git and utilizes its infrastructure for version tracking, the underlying \textit{phylogenetic graph} in EvoGit is fundamentally different from the conventional Git commit graph.
While both structures encode version histories and lineage information, EvoGit introduces semantic constraints and structural interpretations that go beyond typical software development practices.

In standard Git workflows, the commit graph is a general-purpose DAG that captures arbitrary changes to the codebase. 
It imposes no constraints on the nature, scope, or correctness of these changes. 
By contrast, EvoGit enforces domain-specific rules that regulate the evolution of code:
\begin{itemize}
\item \textbf{Partial Order Constraint:} Each descendant version must be no worse than its parent(s), thereby inducing a partial order over the set of versions. This ensures monotonic non-degradation and reflects an evolutionary trajectory of improvement or stability.
\item \textbf{Local Mutation Constraint:} Mutations are restricted to localized edits, typically within a single file or involving the creation of a new, empty module. Large-scale refactoring or disruptive edits across multiple files are explicitly disallowed.
\end{itemize}

These constraints endow EvoGit’s phylogenetic graph with evolutionary semantics, distinguishing it from the unstructured nature of conventional commit histories. In standard development workflows, commits may introduce regressions, remove features, or perform sweeping architectural changes, all of which are permissible in Git, but violate the controlled and incremental design principles enforced by EvoGit.

\section{Discussion on Traceability}

EvoGit ensures full transparency of the evolutionary development process by recording all changes as immutable commits within a Git repository. Each commit represents a distinct, concrete version of the codebase, and every transition (whether introduced by a human or an autonomous agent) is permanently preserved as part of the project's version history.

This immutable and verifiable record enables robust traceability. Any version can be inspected to determine its lineage, identify the precise changes introduced, and analyze the decisions or events that led to its creation. EvoGit enhances this traceability by attaching rich metadata (e.g., build diagnostics, fitness estimates, and agent identifiers) via Git notes, providing a comprehensive account of both the code and its evolutionary context.

Such transparent logging offers benefits beyond reproducibility. It strengthens scientific rigor, supports interpretability, and enables rigorous auditing. In multi-agent systems, where parallel changes and autonomous decisions abound, EvoGit’s traceability allows researchers to reconstruct and analyze the precise contributions and behaviors of each agent. This not only facilitates debugging and performance evaluation, but also offers a foundation for meta-level learning, credit assignment, and trust calibration in future evolutionary frameworks.

\begin{figure}[p]
\centering
\includegraphics[width=0.9\linewidth]{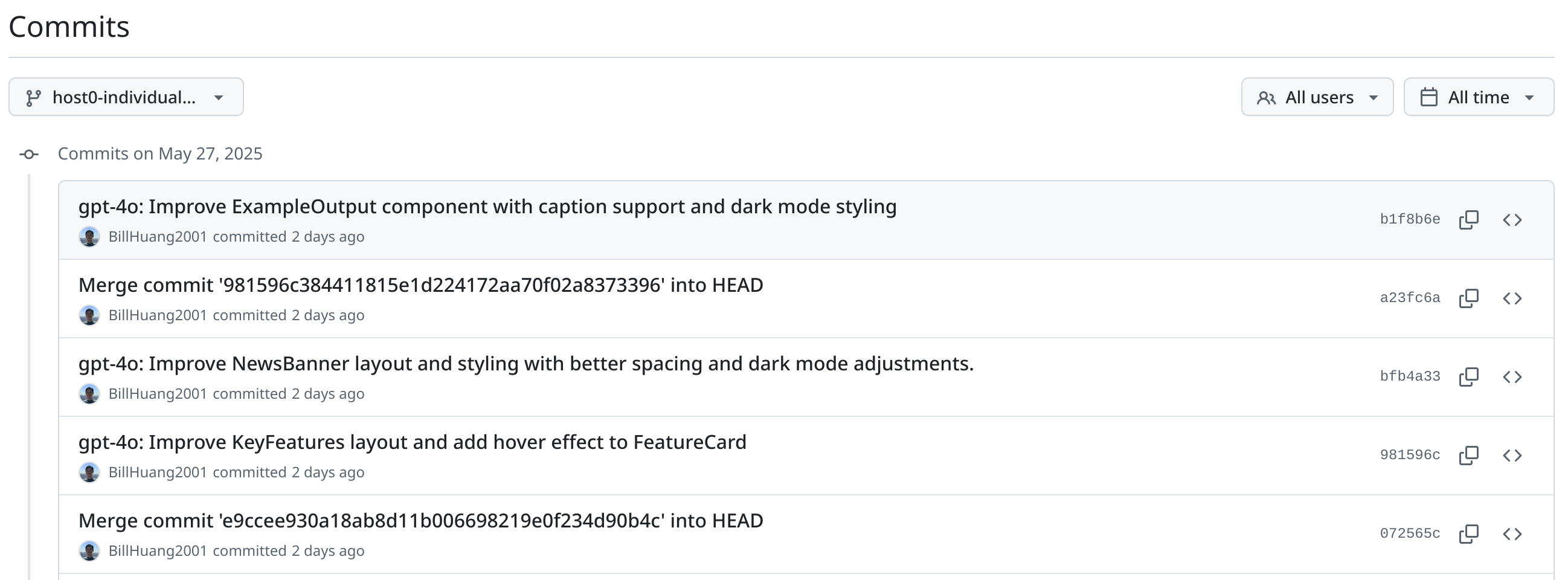}
\caption{Screenshot of the GitHub commit history, showing a sequence of commits generated autonomously by EvoGit agents. Each commit is accompanied by metadata and descriptive messages, enabling chronological inspection of the codebase evolution.}
\label{fig:github-commit-history}
\end{figure}

\begin{figure}[p]
\centering
\includegraphics[width=0.9\linewidth]{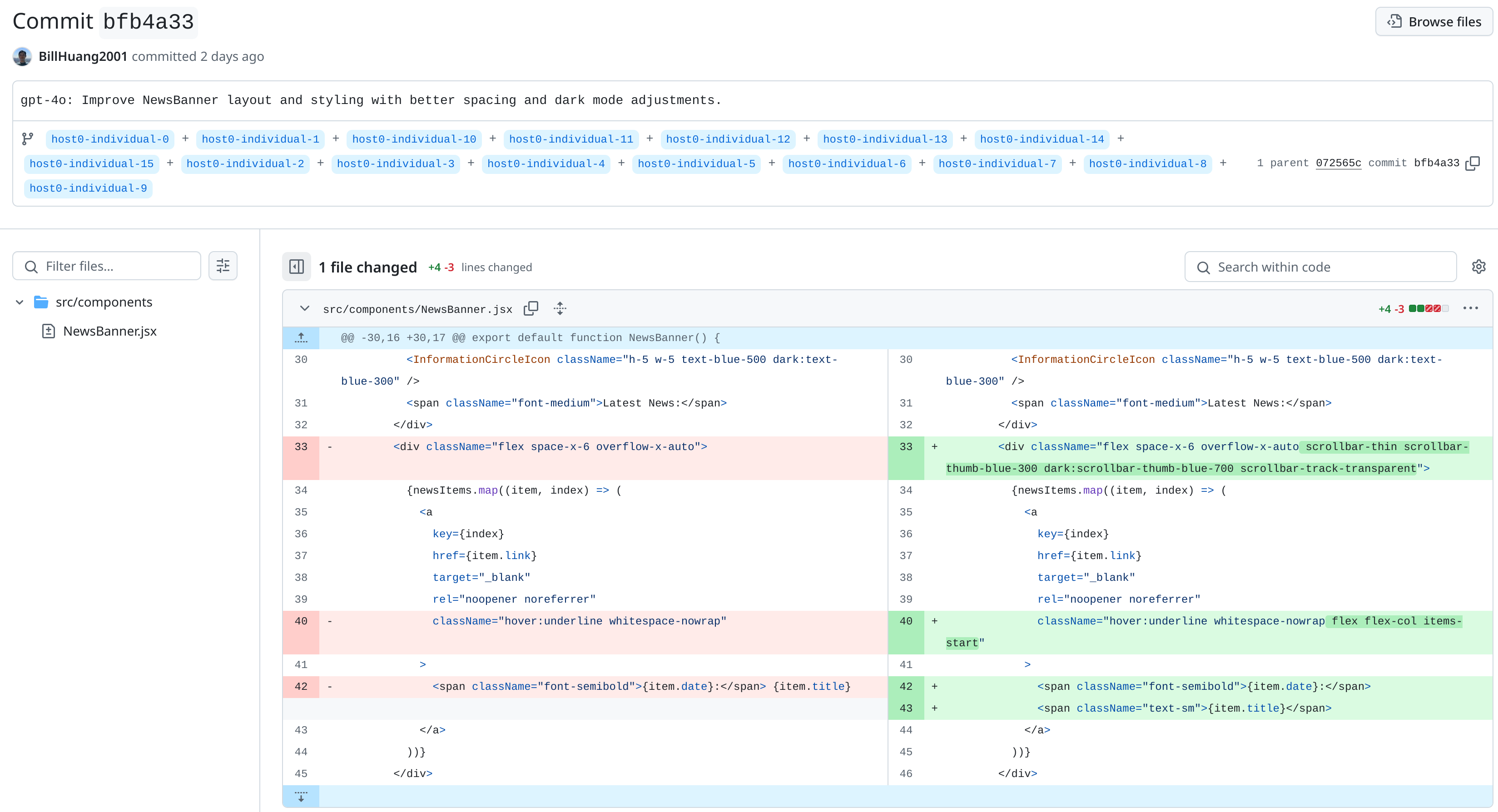}
\caption{Screenshot of GitHub's diff view, illustrating a specific code modification made by an EvoGit agent. The detailed interface supports precise examination of what changed and how the version improved upon its parent. All changes are guaranteed to be minor and localized.}
\label{fig:github-diff-view}
\end{figure}

\begin{figure}[p]
\centering
\includegraphics[width=0.9\linewidth]{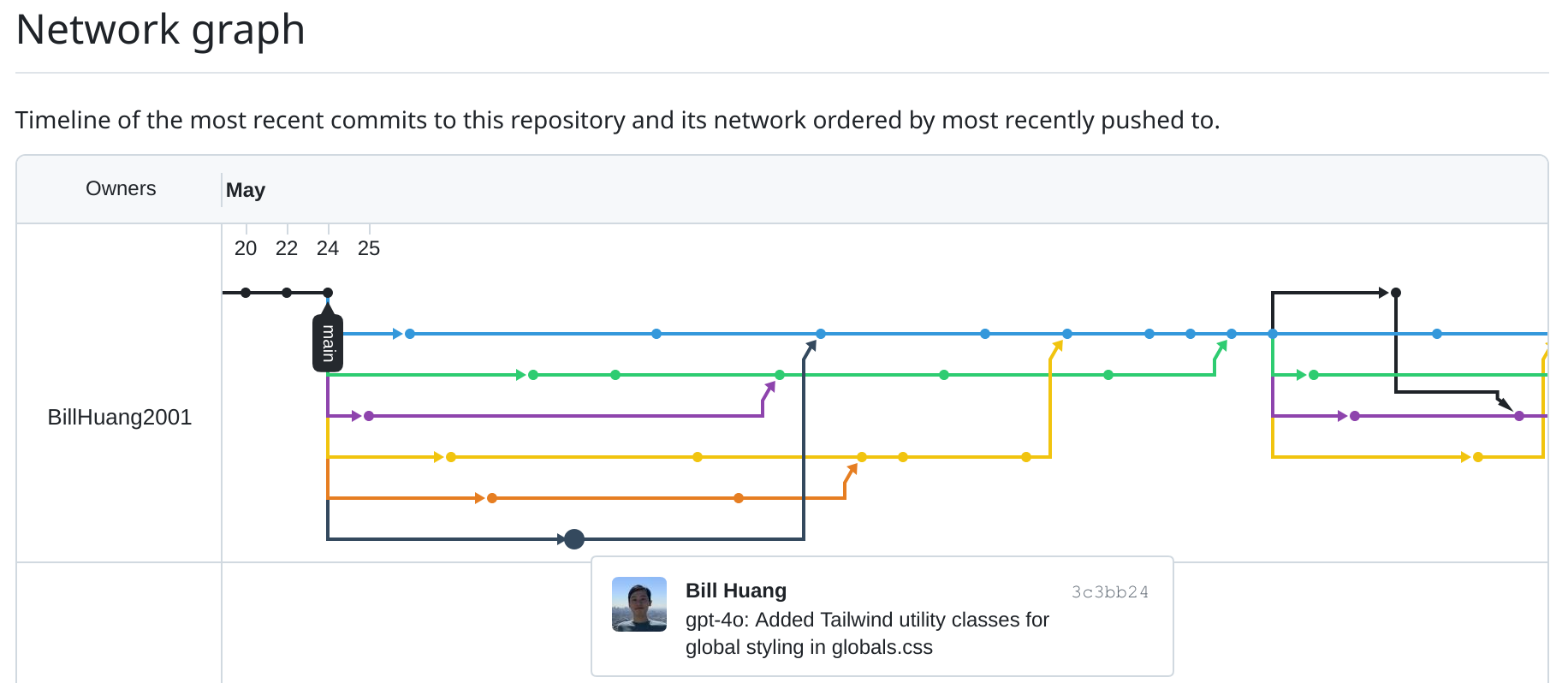}
\caption{Screenshot of the GitHub network graph, visualizing the entire version history as a directed acyclic graph. Each node corresponds to a commit, and branches represent parallel development paths taken by different agents. The \textit{main} branch is initialized by the human; all other branches evolve autonomously. This visualization highlights the decentralized and lineage-preserving nature of EvoGit.}
\label{fig:github-network}
\end{figure}

We demonstrate the traceability of EvoGit in Figs.\ref{fig:github-commit-history},\ref{fig:github-diff-view}, and~\ref{fig:github-network}.

\end{document}